\documentclass[aps,prl,amssymb,secnumarabic,floatfix,superscriptaddress,preprint,nobibnotes]{revtex4-1}
\usepackage{graphicx} 
\usepackage{lineno}
\usepackage{xcolor}
\usepackage[colorlinks=true,citecolor=blue,urlcolor=blue]{hyperref}
\def\textsubscript#1%
{$_{\text{#1}}$}

\def\bluetext#1%
{{\textcolor{red}{#1}}}

\begin{document}

\title{Selective Control of Surface Spin Current in Topological Materials based on Pyrite-type OsX\textsubscript{2} (X = Se, Te) Crystals
}

\author{Yuefeng Yin}
\affiliation{ARC Centre of Excellence in Future Low-Energy Electronics Technologies, Monash University, Australia}
\affiliation{School of Physics and Astronomy, Monash University, Australia}
\affiliation{Department of Materials Science and Engineering, Monash University, Australia}
\author{Michael S. Fuhrer}
\affiliation{ARC Centre of Excellence in Future Low-Energy Electronics Technologies, Monash University, Australia}
\affiliation{School of Physics and Astronomy, Monash University, Australia}
\author{Nikhil V. Medhekar}
\email{nikhil.medhekar@monash.edu}
\affiliation{ARC Centre of Excellence in Future Low-Energy Electronics Technologies, Monash University, Australia}
\affiliation{Department of Materials Science and Engineering, Monash University, Australia}

\date{\today}

\newpage

\begin{abstract}

Topological materials host robust surface states, which could form the basis for future electronic devices. As such states have spins that are locked to the momentum, they are of particular interest for spintronic applications. Understanding spin textures of the surface states of topologically nontrivial materials, and being able to manipulate their polarization, is therefore essential if they are to be utilized in future technologies. Here we use first-principles calculations to show that pyrite-type crystals OsX\textsubscript{2} (X= Se, Te) are a class of topological material that can host surface states with spin polarization that can be either in-plane or out-of-plane. We show that the formation of low-energy states with symmetry-protected energy- and direction-dependent spin textures on the (001) surface of these materials is a consequence of a transformation from a topologically trivial to nontrivial state, induced by spin orbit interactions. The unconventional spin textures of these surface states feature an in-plane to out-of-plane spin polarization transition in the momentum space protected by local symmetries. Moreover, the surface spin direction and magnitude can be selectively filtered in specific energy ranges. Our demonstration of a new class of topological material with controllable spin textures provide a platform for experimentalists to detect and exploit unconventional surface spin textures in future spin-based nanoelectronic devices.


\end{abstract}
\maketitle

\section{Introduction}
Materials with nontrivial topological properties provide a rich playground for discovering unconventional fermions as well as unveiling novel physical phenomena such as giant magnetoresistance and superconductivity \cite{RevModPhys.82.3045,bradlyn2017topological}. Over the last few years, topological states of matter have been revealed in a diverse spectrum of electronic structures from insulators \cite{zhang2009topological} to semimetals \cite{xu2015discovery,liu2014discovery} and metals \cite{bzduvsek2016nodal,PhysRevB.93.121113}. 
Several strategies have been proposed to effectively predict new materials with nontrivial topology by combining the knowledge of dimensionality, crystal symmetry and band theory \cite{bradlyn2016beyond,bradlyn2017topological}. 
In particular, seeking 
materials that can demonstrate highly orientational and controllable spin structures is rapidly emerging as an active area of research \cite{he2018bilinear,fujishiro2018large}. These materials have the potential of exhibiting exotic spin-dependent transport properties, such as inverse spin Hall effect and spin-transfer torque on the surface states 
\cite{PhysRevLett.113.196601,doi:10.1021/am403634u}, which can be beneficial for the development of spintronics and spin detection devices\cite{soumyanarayanan2016emergent,vsmejkal2018topological}.

Each class of topological materials possesses unique spin textures associated with their surface states and spin orbital coupling (SOC)
\cite{PhysRevB.89.155116,PhysRevLett.115.217601}. In topological insulators (TIs), the spin textures of the surface states exhibit a strong spin-momentum locking behavior, where the direction of the spin of a Dirac fermion is locked perpendicular to its momentum \cite{PhysRevLett.107.207602,PhysRevLett.111.066801}, typically lying in the plane of the surface. This character changes significantly for Dirac/Weyl semimetals (DSM/WSM). The spin textures of these materials are altered by the presence of the nodal points, i.e. the crossing points of valence and conduction bands in the bulk. 
In DSM such as Na\textsubscript{3}Bi, the spin polarization tends to vanish at these nodal points due to the recovery of spin degeneracy \cite{PhysRevB.85.195320,xu2015observation}. In WSM, the spin texture is constrained by the presence of local crystalline symmemtries. Near the projection of the Weyl nodes, the spin polarization directions are normally opposite, potentially inferring the chirality of the Weyl nodes. \cite{PhysRevLett.115.217601,PhysRevB.94.195134}. 
In recent years, complicated spin textures have also been observed in metallic surfaces with strong SOC (e.g. silver, tungsten) \cite{ast2007giant,PhysRevB.93.161403} and heterostructures (e.g TI/ferromagnet heterostructures such as Bi\textsubscript{2}Se\textsubscript{3}/NiFe) \cite{wang2017room}. In these cases, electronic spin orientations are heavily affected by the local surface/interfacial structure and can be tuned by doping and adsorption \cite{soumyanarayanan2016emergent}. 
Moreover, external means such as strain and magnetic fields have been explored to achieve a tunable spin texture so that topological materials can be used to fabricate electronic nanodevices \cite{li2018spin}. Despite these promising developments, the search for a topological material with highly energy- and orientation-dependent spin character remains a challenge.
 
Pyrite structure is one of the common crystal structures found in nature \cite{PhysRevB.60.14035}. A few pyrite-type materials have been theoretically predicted to show topologically nontrivial semi-metallic behavior \cite{PhysRevB.91.205128}. The preservation of 3D nodal points under SOC in these materials provides a plausibility for  
 interesting physical phenomena. Recent experiments have confirmed that these topologically nontrivial phases can lead to extremely large magnetoresistance and the emergence of superconductivity \cite{PhysRevLett.118.256601,PhysRevB.96.060509}. Motivated by these findings, here we 
 investigate the family of pyrite-type noble metal selenides and tellurides. Most of these compounds are non-magnetic and reported as either semiconducting with a small band gap or metallic \cite{PhysRevB.88.235208}.
  Among these materials, OsSe\textsubscript{2} and OsTe\textsubscript{2}  are good candidates for exploring the topologically nontrivial electronic structure due to the strong SOC effects of the outermost 5d electrons of Os  \cite{montoya2007n}. Moreover, they have the least magnitude of band gap opening between the conduction and the valence bands compared with other noble metal compounds \cite{PhysRevB.88.235208}. This feature implies a possible band inversion due to inclusion of 
a strong SOC in the vicinity of the Fermi level, an indication of the topologically nontrivial electronic structure. 
Most interestingly, the 5d electrons of Os have shown significant spin anisotropy in oxides due to to a combination of a strong SOC and electronic correlation effects \cite{PhysRevLett.112.147202}. This observation indicates that osmium-based compounds can potentially be used for spin-based devices. 
Finally, some Os-based compounds with Os in high oxidation state (+8) are known to be volatile and toxic. However, the toxicity and environmental impact of  OsSe\textsubscript{2} and OsTe\textsubscript{2} are yet unknown. Earlier work in the literature has shown that the Osmium dioxide, which has Os in the same oxidation state (+4) as OsSe\textsubscript{2} and OsTe\textsubscript{2}, as non-toxic and environmentally benign \cite{smith1974osmium}.

Here we report existence of unconventional  three-dimensional energy- and direction-dependent spin textures in the surface of pyrite-type OsX\textsubscript{2} (X= Se, Te) crystals. OsSe\textsubscript{2} and OsTe\textsubscript{2} are semimetallic with a small energy overlap between conduction and valence bands. The characters of conduction and valence bands are inverted at $\Gamma$ with nontrivial topological indices 1;(000) under strong SOC. The conduction and valence bands are separated everywhere in momentum space by an energy gap. In sharp contrast to the largely in-plane spin texture observed for surface states of other topological insulators, the surface bands of OsSe\textsubscript{2} and OsTe\textsubscript{2} show a spin texture of both in-plane and out-of-plane spin components, with nearly perfect coupling of one momentum direction to the out-of-plane spin component at certain energies protected by complicated local crystal symmetries consisting of $D_{2d}$ and $C_{3v}$. This unconventional spin texture opens new possibilities for injecting or detecting the out-of-plane spin component in topological spintronic devices.

\section{Results and Discussions}

\subsection{Crystal Structure and Bulk Electronic Band Structure}

OsSe\textsubscript{2} and OsTe\textsubscript{2} are both transition metal dichalcogenides crystallized in pyrite-type structure. The single crystals of pyrite-type OsSe\textsubscript{2} and OsTe\textsubscript{2} can be synthesized via chemical vapor transport and are shown to be stable in ambient conditions \cite{Muller1991}. Other crystalline forms of these two materials have not been reported. 
The pyrite structure of OsSe\textsubscript{2} and OsTe\textsubscript{2} is chemically stable and robust against high temperature and pressure \cite{Muller1991,doi:10.1139/v67-437}. 
The crystal structure of OsSe\textsubscript{2} and OsTe\textsubscript{2} belongs to the 
 space group  $Pa\overline{3}$ (SG 205) and is shown  in Figure 1 (a). The Os atoms are located at face-centered-cubic (FCC) sites. Each Se/Te atom is coordinated with three Os atoms and one Se/Te atom, forming a distorted tetrahedron. The lattice parameters optimized using DFT calculations are $a=b=c=$ 6.08 \AA  \ (OsSe\textsubscript{2}) / 6.47 \AA \ (OsTe\textsubscript{2}). These values are close to the values reported by previous experimental investigations \cite{stassen1968crystal,Muller1991}.

First we study the electronic properties of OsSe\textsubscript{2}/OsTe\textsubscript{2} by examining the projected density of states (PDOS) and the electronic band structure in the absence of SOC. In case of OsSe\textsubscript{2} (see Figure 2 (a)), the states near the Fermi level are mainly from Os $d$ states and Se $p$ states. The Os $d$ states are dominant below the Fermi level, while Se $p$ states dominate from the Fermi level up to 1.07 eV above the Fermi level (inset graph of Figure 2 (a)). 
These features are well reflected in the band structure diagram shown in Figure 2 (b). The band structure of OsSe\textsubscript{2} shows a semimetallic character, with the conduction band near $\Gamma$ partially filled, accompanied by a hole pocket along $\Gamma$ - Y in the valence band. We further divide the contribution from Os $d$ orbitals split into Os $d$ $t_{2g}$ and $d$ $e_{g}$ states due to the ligand field splitting effect. We find that the partially-filled low-energy conduction band across the Fermi level near $\Gamma$ is parabolic and mainly has Se $p$ states. The adjacent valence bands around $\Gamma$ are predominantly from Os $d$ $e_{g}$ orbitals, whereas other low-lying valence bands below the Fermi level are of largely Os $d$ $t_{2g}$ character. We also note the existence of multiple degeneracy points below the Fermi level at high symmetry points or along high symmetry lines. These band structure characteristics are consistent with previous investigations on other pyrite-type chalcogenide compounds (e.g. FeS\textsubscript{2}, RuSe\textsubscript{2}) \cite{PhysRevB.88.235208}. 

Upon inclusion of spin orbit coupling (SOC), the electronic states near $\Gamma$ point change significantly as presented in Figure 2 (c). The bottom of the partially occupied conduction band is now occupied by Os $d$ states and becomes flat near $\Gamma$ point and is fully gapped from the adjacent valence band (see Supplementary Note C and Supplementary Figure S6 for details). The Se $p$ states are now pushed under the fourfold degenerate highest occupied valence bands (considering the spin), around -0.22 eV below the Fermi level. The Os $d$ states are pushed up to not just the bottom flat section of the conduction band, they are also found in adjacent regions up to the Fermi level. This change in band ordering is different from the conventional band inversion in that the band inversion in this case occurs between adjacent bands. This can be explained by the energy pinning of the highest occupied valence bands. We note that without SOC effects, the highest occupied bands are sixfold degenerate at $\Gamma$ point. This degeneracy becomes fourfold upon consideration of SOC and is resulted from crossing of bands with different pairs of $C_{3}$ eigenvalues. The fourfold degenerate bands consist of two bands with $C_{3}$ eigenvalues of $\pm e^{i\pi/3}$ and two bands with $C_{3}$ eigenvalues of -1. The energy position of this band degeneracy does not move with SOC, therefore the inverted Se $p$ band is located further below the fourfold degenerate bands. This band inversion feature is a strong signal of the existence of nontrivial topological phases in OsSe\textsubscript{2}. 
It should be emphasized that since GGA functional is often known to underestimate the energy alignment between conduction bands and valence bands, we employed hybrid functional HSE06  to confirm the existence of the non-trivial band structure \cite{doi:10.1063/1.1564060}. 
Our HSE06 calculations show that 
the band inversion between Os $d$ states and X (X= Se, Te) $p$ states is retained (see Supplementary Note A and Supplementary Figure S2). Therefore we can conclude that the band order of OsX\textsubscript{2} possesses nontrivial features.

SOC also has a significant impact on band crossings at high symmetry points and high symmetry lines in momentum space. Among all band crossings in the band structure diagram observed without SOC, most of them are gapped with SOC turned on owing to band repulsion. However, along the $C_{3}$-invariant $\Gamma$ -- R line, some band crossings can survive the SOC effects. We also observe $C_{3}$-protected sixfold degenerate bands at R point. These features agree well with previous theoretical predictions that pyrite-type materials can have six-fold fermions stabilized by space group symmetries \cite{bradlyn2016beyond,FURUSAKI2017788}. Moreover, the band dispersion along Y -- M forms four-fold degenerate nodal lines with the presence of two-fold screw rotations about the x, y and z axes $\widetilde{C}_{2\alpha}$, while along X -- M the bands are still two-fold degenerate (Figure 2 (d). This anisotropy behavior suggests that nodal points can exist under strong SOC effects with the protection of nonsymmorphic crystal symmetries such as screw rotation \cite{FURUSAKI2017788} (See a detailed discussion on the existence of symmetry-protected nodal lines in Supplementary Note D, Supplementary Figure S7 and Supplementary Table S1).

Figure 2 (e) and (f) present the band structure diagram for OsTe\textsubscript{2} with and without SOC effects using the same approach for OsSe\textsubscript{2}. The orbital composition of conduction and valence bands is similar to that observed in OsSe\textsubscript{2}. The electronic structure of OsTe\textsubscript{2} is also semi-metallic (the unfilled hole packet is found along $\Gamma$ - M. 
A key difference between OsSe\textsubscript{2} and OsTe\textsubscript{2} without SOC is that there is an accidental Dirac-like touching point between the low-energy conduction band of Te $p$ states and adjacent valence bands of Os $d$ states. We also find degeneracy of bands near $\Gamma$ and R point consisted of Te $p$ states at about 1.1 eV below the Fermi level. This is in contrast to clear separation of Os $d$ states and Se $p$ states in OsSe\textsubscript{2}. The difference in the electronic structure between OsSe\textsubscript{2} and OsTe\textsubscript{2} can be attributed to the bonding between chalcogen dimers. The Te-Te covalent bond distance (2.87 \AA) in OsTe\textsubscript{2} is much larger than Se-Se bond distance (2.57 \AA) in OsSe\textsubscript{2}. The increase in the bond distance tends to downshift the conduction band minimum and affect the shape of the band. These results indicate that the electronic structure of these chalcogenides can be tuned dynamically via manipulating internal atomic coupling. 
The band structure of OsTe\textsubscript{2} also undergoes significant changes with SOC (Figure 2 (f)). The conduction band minimum near $\Gamma$ becomes flat and is of Os $d$ $e_{g}$ character. The Te $p$ states are inverted inside the valence bands. Degeneracy points are found along the $\Gamma$ -- R line. All these features qualitatively agree with the electronic structure OsSe\textsubscript{2}, especially in the low energy region, therefore in the remaining part of the paper we will focus on OsSe\textsubscript{2} (results on OsTe\textsubscript{2} can be found in Supplementary Figures S12 and S13).

We next calculate the  $Z_{2}$ topological invariants based on the evolution of Wannier charge centers to confirm the existence of topological nontrivial phases in OsSe\textsubscript{2} and OsTe\textsubscript{2} \cite{PhysRevB.83.035108}. 
The $Z_{2}$ topological invariants for the 3D bulk OsSe\textsubscript{2}/OsTe\textsubscript{2} crystal $\nu_{0}$;($\nu_{1}\nu_{2}\nu_{3}$) are obtained by tracing the evolution of Wannier charge centers of fully occupied Bloch bands for six time-reversal invariant momentum planes ($k_{1}=0$ and $\pi$, $k_{2}=0$ and $\pi$, $k_{3}=0$ and $\pi$). To demonstrate this, we show the evolution of Wannier charge centers along $k_{2}$ for planes $k_{3}=0$ and $\pi$ for OsSe\textsubscript{2} with SOC as an example (Figure 3). Results for other planes can be found in Supplementary Figure S9. It can be clearly seen that $Z_{2}=1$  for $k_{3}=0$ plane since the reference line has odd number of intersections with the evolution lines (Figure 3 (a)),  whereas even number of crossings between the reference line and evolution lines indicates that $Z_{2}=0$  for  $k_{3}=\pi$ plane (Figure 3 (b)). We find the $Z_{2}$ indices for both OsSe\textsubscript{2} and OsTe\textsubscript{2} are 1; (000), proving that these pyrite crystals are topologically nontrivial.

Our results show that the lowest conduction band and neighboring valence bands for OsSe\textsubscript{2} and OsTe\textsubscript{2} are completely gapped along high symmetry lines, providing a possible venue for the emergence of nontrivial surface states between them. Previous investigations have implied that these states can be linked with novel physical properties such as giant magnetoresistance\cite{PhysRevLett.118.256601} and superconductivity\cite{PhysRevB.96.060509}. Another interesting electronic feature is the formation of inverted flat band at the bottom of the conduction band near $\Gamma$. This is similar to the profile of the well-known topological Kondo insulators such as SmB\textsubscript{6} \cite{li2014two}. Currently the flat band has only been realized in materials with strong correlated $f$ electrons through hybridization \cite{neupane2013surface}. Therefore the stability of this flat band in OsSe\textsubscript{2}/OsTe\textsubscript{2} consisting of hybridized Os $d$ states may need to be further assessed with advanced computational schemes \cite{kang2015band}.

\subsection{Surface Electronic Structure}

Next we discuss the surface correspondence of the bulk nontrivial electronic structure in detail. The morphology and electronic structures of the surfaces of pyrite crystals have been well studied before by both experiments and theoretical calculations\cite{doi:10.1021/jp0009498,doi:10.1021/jp100578n,MURPHY20091}. The (001), (110), (111) and (210) surfaces have all been reported as possible cleavage surfaces in pyrite systems \cite{doi:10.1021/jp100578n}. The rich symmetry of the pyrite crystal also allows the existence of multiple surface terminations for each cleavage surface \cite{doi:10.1021/jp100578n}. Here we selected the appropriate surface to investigate based on two important factors: thermodynamic stability and extent of surface reconstruction \cite{doi:10.1139/v67-437}. The (001) surface is known to be the most common cleavage surface in pyrite-type crystals and  thermodynamically the most stable. The most stable termination of (001) surface can well retain the bulk-like atomic configurations with little surface reconstruction \cite{doi:10.1021/jp100578n}. The (110) surface is much less stable than the (001) surface, but the atom displacement at the surface is small. The (111) and (210) surfaces can be more stable than (110), but at the cost of significant surface reconstructions to compensate for the loss of coordination on the surface. Due to the considerable computational cost in obtaining optimal reconstructed surfaces involving heavy Os atoms, 
here we mainly consider the (001) surface and briefly discuss the characteristics of the (110) surface. . 
Given the symmetry of the crystal, the (001) surface of OsSe\textsubscript{2} can have three possible surface terminations: Se-Se terminated, Se terminated and Os terminated (see Supplementary Figure S10). Our slab cell calculations and previous investigations have shown that the Se-Se terminated and Os-terminated surfaces are less energetically favorable compared to Se-terminated surface \cite{doi:10.1021/jp0009498,doi:10.1021/jp100578n}. In addition, Se-Se terminated and Os-terminated surfaces have significant surface reconstructions upon relaxation. In the following, we will focus on the electronic structure of Se-terminated (001) surface.

The relaxed Se-terminated (001) surface OsSe\textsubscript{2} and OsTe\textsubscript{2} (see Figure 4 (a)) retains the bulk atomic configurations,  consistent with earlier theoretical predictions on other pyrite-type crystals \cite{doi:10.1021/jp100578n}.  The Se-terminated (001) surface preserves the surface stoichiometry and only features rupture of Os-Se bonds normal to the (001) plane upon cleavage (i.e. no Se-Se bonds are broken). As a consequence the coordination of surface Os atoms changes significantly. In the pristine lattice, each Os atom is surrounded by six Se atoms, forming a distorted octahedral (point group $O_{h}$). On the other hand, the coordination of each Os on (001) surface atom changes to square-planar pyramidal (point group $C_{4v}$) due to the loss of one Se ligand. The loss of coordination normally leads to further splitting of Os $d$ states (between $d_{x^{2}-y^{2}}$ and $d_{z^{2}}$) and formation of spin polarized surface states, as indicated in previous reports \cite{doi:10.1139/v67-437}.

To effectively obtain the topologically nontrivial surface electronic structures of OsSe\textsubscript{2} and OsTe\textsubscript{2},  we perform surface state calculations by the Green's function approach using a tight-binding model Hamiltonian derived from our first principles calculations \cite{sancho1985highly}. As seen from Figure 4 (b,c), we see different band dispersions along the lines connecting $\overline{X}$ (0.5,0) with $\Gamma$ (0,0) ($\overline{M}$ (0.5,0.5)) and lines connecting $\overline{Y}$ (0,0.5) with $\overline{\Gamma}$ ($\overline{M}$). (i.e. $\overline{X}$ and $\overline{Y}$ are not equivalent.) This reflects the anisotropy observed in the bulk band structure. Moreover, the lowering symmetry of surface Os atoms results in a complicated electronic structure at the surface under the influence of strong SOC interactions near the Fermi level. In the vicinity of the Fermi level (ranging from -0.03 eV to -0.09 eV), we can observe hole-like surface bands around $\Gamma$ point (denoted by S1 ($\overline{\Gamma}$ -- $\overline{M}$), S2 ($\overline{\Gamma}$ -- $\overline{X}$) and S3 ($\overline{\Gamma}$ -- $\overline{Y}$) in Figure 4). These surface bands are partly buried in the partially occupied bulk conduction bands and connect with the underlying bulk valence bands.

The anisotropy in the surface electronic structure prompts us to further examine the spin structure by analyzing spin components $S_{x}$, $S_{x}$ ad $S_{z}$ along surface band S1, S2 and S3 in Figure 4 (d). We find that each surface band shows a distinctive energy- and orientation- dependent spin texture profile. 
For the S1 band (along $\overline{\Gamma}$ -- $\overline{M}$), we observe 
a strong $S_{z}$ polarization near the Fermi level ($\sim$0.04 eV below the Fermi level). For energies deeper below the Fermi level, the $S_{y}$ component of the S1 band tends to decrease and the magnitude of the $S_{x}$ component gradually increases. At $E_{F} - 0.08 $ eV, the $S_{y}$ component of the S1 band reduces to only 10\% of the $S_{x}$ component. 
For the S2 band (along $\overline{\Gamma}$ -- $\overline{X}$), the surface spins are composed of little in-plane spin component and a large out-of-plane spin component. The total magnitude of the spin reduces significantly as the energy level decreases. 
Finally, the S3 band (along $\overline{\Gamma}$ -- $\overline{Y}$) has a strong spin polarization along the x axis and almost no spin component in other two directions. These distinct surface spin textures shown in Figure 4 (d) indicate the presence of topological nontrivial phases in OsSe\textsubscript{2}.

We also used slab calculations based on first principles DFT methods to verify the unusual surface spin textures shown in Figure 4 (d) 
near the Fermi level as the tight binding model usually leads to small difference in energy alignment when compared to DFT results. 
As shown in Figure 4 (e), the slab calculation reveals that the $d$ orbital electronic states from top layer Os atoms are the dominant contributors to the surface bands observed in Figure 4 (b,c). The orbital character of the surface bands shows a further splitting of Os $t_{2g}$ and Os $e_{g}$ states as well as change of energy ordering compared to the bulk electronic states (see Fig. S3). The S1 band ($\overline{\Gamma}$ -- $\overline{M}$) is mainly comprised of a near equal fraction of $d_{yz}$ and $d_{xz}$ states, followed by $d_{x^{2}-y^{2}}$. The orbital contributions from $d_{z^{2}}$ and $d_{xy}$ are much weaker compared to other $d$ states. The S2 band ($\overline{\Gamma}$ -- $\overline{X}$) shows a slight change in the ordering. $d_{yz}$ and $d_{xz}$ are still the strongest d characters in the surface band, while $d_{z^{2}}$ becomes the next strong d states, surpassing $d_{x^{2}-y^{2}}$. In the S3 band (along $\overline{\Gamma}$ -- $\overline{Y}$) $d_{x^{2}-y^{2}}$ states become the second most contribution following $d_{yz}$, while the projection from $d_{xz}$ orbital decreases rapidly. The $d_{z^{2}}$ character becomes negilible. The contribution from $d_{xy}$ is weak in all surface bands.
This anisotropic behavior in orbital splitting can be used to explain the strong direction-dependent spin split in the surface bands as shown in Figure 4 (e). The S1, S2 and S3 bands are separated from their spin splitted counterpart in the bulk by $\sim$90 meV, 120 meV and 30 meV, respectively. The out-of-plane orbital character features of the S2 band ($d_{yz}$, $d_{xz}$ and $d_{z^{2}}$) is a key factor for obtaining such a sizable splitting effect. The significant in-plane $d_{x^{2}-y^{2}}$ character leads to a small band splitting in the S3 band. We note that similar spin splitting behavior has been realized in layered materials with doped surface where the surface symmetry is broken by dopants \cite{PhysRevB.96.085403}. Here we find that symmetry breaking at the pristine pyrite surface can achieve the same effect by inducing different orbital screening along each high symmetry line. This observation indicates a new approach of spin control at the surface from a spintronics viewpoint. Overall, the slab calculations results reaffirm the key electronic features discovered in tight binding models in Figure 4 (b--d).
 
The complex surface spin texture can be visualized by a three-dimensional perspective view of the constant energy contour at  0.07 eV below the Fermi level in Figure 5 (a). It is evident that the spin direction and magnitude vary in the momentum space and the transition from out-of-plane spin polarization to in-plane polarization can be clearly observed. Figure 5 (b) shows the evolution of the in-plane spin character by plotting the spin texture in the iso-energy surface spectral function projected on to the 2D Brillouin zone using the Wannier tight-binding Hamiltonian. We select three representative energies: 0.05 eV, 0.07 eV and 0.09 eV below the Fermi level. The surface bands are projected in the iso-energy surface spectral function map of $E_{F}-0.05$ eV and $E_{F}-0.07$ eV as a ring-like contour surrounding the bulk states.  At $E_{F}-0.05$ eV, the spin texture shows the helical spin texture similar to that observed in topological insulators around the $\overline{\Gamma}$ -- $\overline{Y}$ axis, where the $S_{x}$+$S_{y}$ spin orientation is tangential to the iso-energy surface contour (i.e. spin-momentum locking). The magnitude of the in-plane spin component gradually decreases from $\overline{\Gamma}$ -- $\overline{Y}$ axis to the $\overline{\Gamma}$ -- $\overline{X}$ axis, and finally vanishes at the the $(\pm k_{x},0)$ point. For the out-of-plane spin component ($S_{z}$), the magnitude is nearly zero at the intersection with the $\overline{\Gamma}$ -- $\overline{Y}$ axis. The trend of the magnitude of $S_{z}$ along the surface contour is opposite to the in-planar spin component. When the in-plane spins disappear, the out-out-plane spins reach maximum. The spin texture shows a slightly different behavior at $E_{F}-0.07$ eV, with the in-plane spins no longer tangential to the surface contour. As the energy reaches $E_{F}-0.09$ eV, parts of the surface bands are now touching the bulk and the contour breaks into linear segments. The in plane spin texture on these segments is mostly $S_{x}$ polarized, with the out-of-plane texture shows a strong positive $S_{z}$ polarization. This kind of highly energy- and direction-dependent surface spin texture is yet to be observed in topologically nontrivial materials. 

To effectively demonstrate the distinct spin behavior of the surface band structure, we next construct a three-dimensional view of the surface cone in an energy range of [-0.10, 0] eV below the Fermi level (see Figure 5 (c)).
The shape of the surface contour is significantly reshaped by intersecting with the bulk states. Along the x axis in the range of [-0.035, 0] eV below the Fermi level, we can see a near-rectangular valley cut in the cone (denoted by black dashed lines). 
The width of the cut agrees well with the projection of the bulk flat conduction band. However, this cut is not observed along the y axis in the same energy range. Another special feature in the cone is the arc cut along the y axis near 0.08 eV below the Fermi level (denoted by the green dashed line). This surface cut eliminates the surface states near $k_{y}=0$ starting from 0.082 eV below the Fermi level. A similar small cut is found along the x axis near 0.09 eV below the Fermi level. This means another asymmetrical surface dispersion between 0.082 eV and 0.09 eV below the Fermi level. This feature is opposite to the signal on the upper side of the cone as it screens out the $k_{y}=0$ states instead of $k_{x}=0$ states. 
These results indicate a possible way of selectively modulating spin texture from in-plane (near $\overline{Y}$)  to out-of-plane (near $\overline{X}$) by a combination of energy and momentum direction.

Figure 5(d) shows the magnitudes of spin polarizations $S_{x}$, $S_{y}$ and $S_{z}$. The circular and arc-like contour from inside to outside in Figure 5 (d) refer to the iso-energy surface spectral function at $E_{F}- $0.03/0.04/0.05/0.06/0.07/0.08/0.09/0.095 eV. The opposing colors in $S_{x}$, $S_{y}$ and $S_{z}$ graphs indicate the preservation of TR symmetry in the crystal. The spin flips its sign when comparing states at $k$ and $-k$. The total spin polarization gradually decreases as the surface states approach the bulk valence band.

To understand the physical origin of the anisotropic spin polarization on the surface and why it is different from conventional topological materials, we conduct a symmetry-based analysis using a $k\cdot p$ Hamiltonian (see Supplementary Note B and Supplementary Figure S3, S4 and S5 for details). Our results show that the (001) pyrite surface possesses a much more complicated local symmetry than the expected $C_{4v}$ symmetry. The small structural distortion places the surface Os atom in a local bonding environment with a mixing of $D_{2d}$ and $C_{3v}$ symmetries. Both these symmetries can be regarded as further symmetry reduction of $C_{4v}$ symmetry via loss of coordination. Previous reports have indicated that crystals with $C_{3v}$ symmetry can exhibit large out-of-plane spin polarization, while $D_{2d}$  symmetry leads to mainly in-plane spin components\cite{PhysRevB.97.085433,PhysRevLett.103.266801}. We find that $\overline{\Gamma}$ -- $\overline{M}$ and $\overline{\Gamma}$ -- $\overline{X}$ directions have a significant $C_{3v}$ element, while  the band is composed of both $C_{3v}$ and $D_{2d}$ elements along $\overline{\Gamma}$ -- $\overline{Y}$ direction. We confirm that the stability of these surface states is protected by these local symmetries. The anisotropic spin polarization behavior on the surface is also robust against structural perturbations such as point vacancies (see Supplementary Note E and Supplementary Figure S8).

The above analysis on the electronic structure of OsSe\textsubscript{2} (001) surface has already shown promising potentials for OsSe\textsubscript{2} as a topological nontrivial material. We also calculate surface electronic structure of (110) surface (See Supplementary Figure S11). The (110) surface is less stable than (001) surface and is not a common cleavage surface observed in pyrite system. The surface morphology of (110) surface is much complicated than (001) surface. The (110) surface can be either (Os+Se)-terminated or Se-terminated. The biggest difference between (001) and (110) surfaces is that visible surface states can be found near and slightly above the Fermi level in pairs, possessing opposite spin textures. In (001) surface, only positive $S_{z}$ polarized surface bands (S1 -- S3) are found, while the other spin split bands are buried in the bulk. Another difference observed in (110) surface is that the surface states below the Fermi level are largely merged with the bulk states.

 \subsection{Surface spin textures in OsTe\textsubscript{2}}

We find that OsTe\textsubscript{2} shares most features with that found in OsSe\textsubscript{2}, with differences in the energy alignment (see Supplementary Figure S12). The surface bands S1 -- S3 connecting the bulk valence bands and bulk conduction bands are found in the vicinity of the Fermi level. The Surface bands also possess highly anisotropy behavior. The S1 band ($\overline{X}$ -- $\overline{\Gamma}$) is the most visible band, while the S3 band ($\overline{Y}$ -- $\overline{\Gamma}$) is significantly overshadowed by bulk bands. The spin anisotropy behavior is also seen in these surface bands, with the in-plane spin polarization dominating near $\overline{Y}$ and out-of-plane spin polarization prevailing near $\overline{X}$. These results agree qualitatively well with those observed in OsSe\textsubscript{2}. Compared to OsSe\textsubscript{2}, the bulk screening effect on the surface bands is more significant in OsTe\textsubscript{2}. The full ring-like iso-energy contour only exists in the range of $E_{F}-0.03$ eV to $E_{F}-0.06$ eV, only half of that in OsSe\textsubscript{2}. Further below this energy range, the surface contours become segments linking the bulk counterparts (see Supplementary Figure S12 and S13). This suggests that OsTe\textsubscript{2} could also be a feasible candidate for selectively moldulating the nontrivial surface spin structure.

Finally, we discuss the importance of the highly anisotropic surface spin textures observed in OsX\textsubscript{2} (X= Se, Te) from our DFT calculations in relation to current challenges in experimental and theoretical studies on spintronics. Generating and manipulating out-of-plane electronic spins without applying an external electric/magnetic field has been a key challenge in applications such as spin logic devices, which requires both in-plane and out-of-plane spin components to effectively transport spin information. Most previous reports have proposed the anomalous spin Hall effect (ASHE) in ferromagnetic compounds as a viable solution to this challenge since materials with ASHE have spin currents oriented  with respect to the ferromagnet's magnetization direction \cite{chang2013experimental,suzuki2016large}. However, precise control over the spin injection in these magnetic compounds is often difficult to achieve due to their complex structures and vulnerability to structural inhomogeneities at the interface \cite{wang2018direct}. Recently, some reports have offered an alternative route by seeking hidden spin polarizations in non-magnetic solids based on the interplay between bulk crystalline symmetry and atomic site symmetries \cite{zhang2014hidden}. Our calculations have expanded this mechanism by demonstrating that highly directional- and energetic-dependent spin textures can exist on the surface of a bulk inversion-symmetric crystal, protected by local crystalline symmetries. In experiments, this effect can be tested using thin films fabricated by atomic layer deposition technique and the spin polarization behavior could be characterized by spin and angle resolved photoemission spectroscopy (ARPES)\cite{doi:10.1002/chem.201803327}. Moreover, further theoretical investigations can be done on enhancing the tunability of the spin textures in the pyrite crystals, as well as generalizing the theory of finding hidden anisotropic spin polarizations in covalently-bonded layered structures. We expect that the physical insights obtained from these studies could be beneficial for developing next-generation spintronic transistor and sensing devices.

 In summary, using density functional theory calculations, we show highly anisotropic spin textures near the Fermi level on the low energy surfaces of topological nontrivial compounds OsX\textsubscript{2} (X = Se, Te). The electronic structure of these pyrite-type crystals in the bulk form is nonmagnetic and semi-metallic. The low-energy energy bands near $\Gamma$ show significant band inversion due to SOC interactions. These states are found to be topologically nontrivial and stable owing to the presence of crystal symmetries. For stable cleavage surface (001) of these crystals, we observe exotic spin-anisotropic surface bands connecting the partially occupied conduction and valence bands. 
These surface bands have negligible in-plane spin polarization near $\overline{X}$, while around $\overline{Y}$ the in-plane spin components become dominant. The spin texture evolution of the surface bands is heavily influenced due to the energy screening by bulk bands. These anisotropic features are protected by the presence of local $C_{3v}$ crystal symmetry. 
We anticipate that the anisotropic surface spin textures in topologically nontrivial pyrite crystals as predicted here could be verified by the ARPES measurements. These measurements can also provide experimental evidence for the exotic surface states associated with the bulk nodal line structures. 
Such unconventional energy- and direction-dependent spin texture could be beneficial for potential spintronics applications. Moreover, discovery of these physical phenomena on a pristine surface of a strong topological material could inspire new strategies of searching for exotic topological nontrivial properties. 

\section{Methods}
The optimized geometry and the electronic structure of OsSe\textsubscript{2}/OsTe\textsubscript{2} are obtained using density functional theory (DFT) as implemented in the Vienna ab initio Simulation Package (VASP) \cite{PhysRevB.54.11169}. The Perdew-Burke-Ernzehof (PBE) form of the generalized gradient approximation (GGA) is used to describe electron exchange and correlation \cite{PhysRevLett.77.3865}. 
We also considered the effect of correlations via the on-site Hubbard U term\cite{PhysRevB.57.1505}, however, the electronic structures show a negligible effect except for small energy shifts in the band alignment (see Supplementary Note A and Supplementary Figure S1). 
The Heyd-Scuseria-Ernzerhof (HSE06) hybrid potential is employed to check for possible overestimation of band inversion within GGA \cite{doi:10.1063/1.1564060}. The kinetic energy cutoff for the plane-wave basis set is set to 400 eV. The standard PBE pseudopotential is adopted in all calculations, treating eight valence electrons for Os ($5d^{6}6s^{2}$) and six valence electrons for Se/Te (Se: $4s^{2}4p^{4}$, Te: $5s^{2}5p^{4}$). We use a  $9\times9\times9$ $\Gamma$-centered k-point mesh for sampling the Brillouin zone. All structures are fully relaxed until the ionic forces are smaller than 0.01 eV/\AA. The surface states spectrum is calculated using the open-source code WannierTools, based on the Wannier tight-binding Hamiltonian obtained from wannier90 \cite{WU2018405,MOSTOFI20142309}. Os $d$ and Se/Te $p$ orbitals are used as initial projectors for tight-binding Hamiltonian construction. 
To confirm the surface states obtained by tight-binding projections, we also explicitly calculated the surface electronic structure using a slab model with a $1\times1\times8$ supercell. A  20 \AA \ vacuum separates the periodic image to avoid any spurious interactions. The surfaces are also fully relaxed with the energy convergence up to $10^{-7}$ eV and a force tolerance up to 0.01 eV/\AA.

\section{Data Availability Statement}

The data and codes that support the findings of this study are available from the authors upon reasonable request.

\section{Acknowledgements}

The authors would like to thank the support from the ARC Centre of Excellence in Future Low-Energy Electronics (CE170100039). M.F. acknowledges the support from Australian Laureate Fellowships (FL120100038). The authors also gratefully acknowledge computational support from the Monash Campus Cluster, NCI computational facility and Pawsey Supercomputing Facility.

\section{Competing Interests}

The authors declare no competing interests.

\section{Author Contributions}

Y.Y. and N.V.M. conceived the work and were responsible for the overall research planning and direction. Y.Y did the first principles calculations and analysis. Y.Y and N.V.M wrote the main text. Y.Y wrote the Supplementary Information and plot all figures and tables. M.S.F. contributed to the revision of the manuscript. All authors contributed to the discussion.  

\section{Supplementary Information}
 
Supplementary Information accompanying the paper is available at npj Quantum Materials website.

\newpage


\newpage

\section{Figures}

 \begin{figure}[htbp]
 \begin{center}
 \includegraphics[scale=0.08]{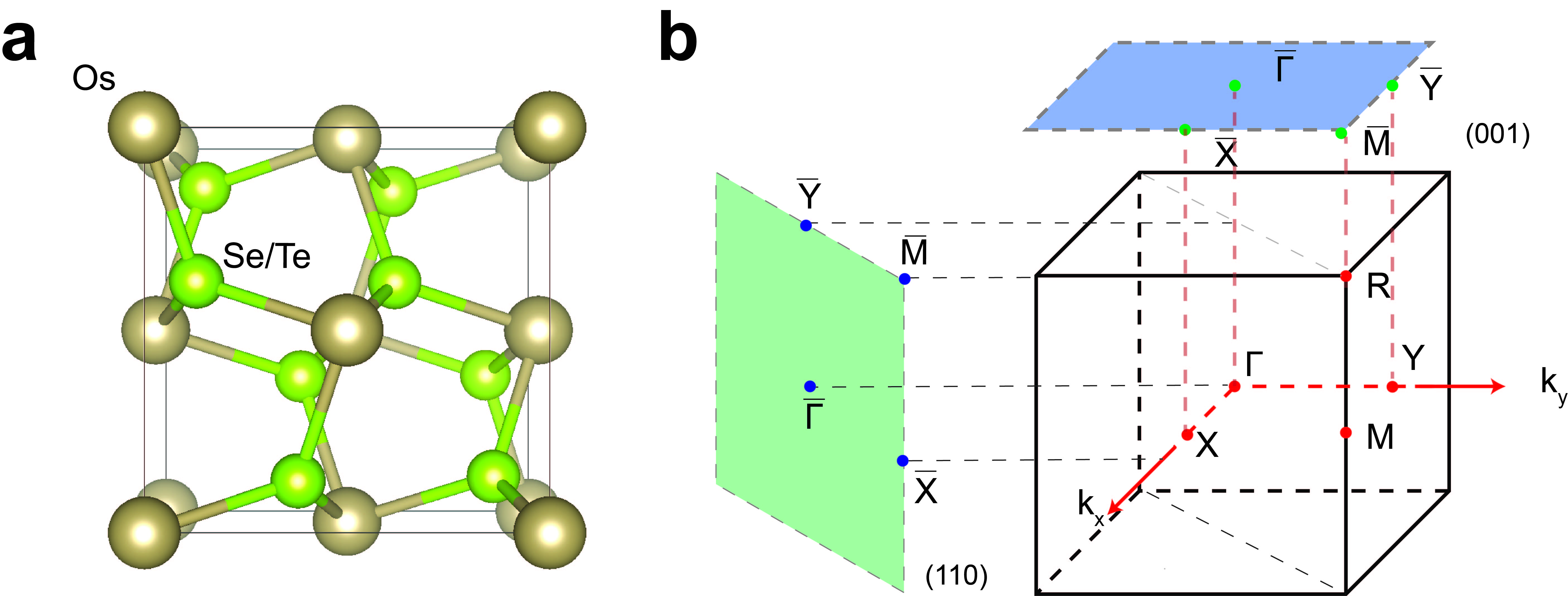}
 \end{center}
 \caption {\small (a) Crystal structure of OsSe\textsubscript{2}/OsTe\textsubscript{2}. (b) The bulk Brillouin zone and the projected surface Brillouin zones for (001) and (110) surfaces.}\label{fig1}
 \vspace{\baselineskip}
 \end{figure}

 \begin{figure}[htbp]
 \begin{center}
 \includegraphics[scale=0.20]{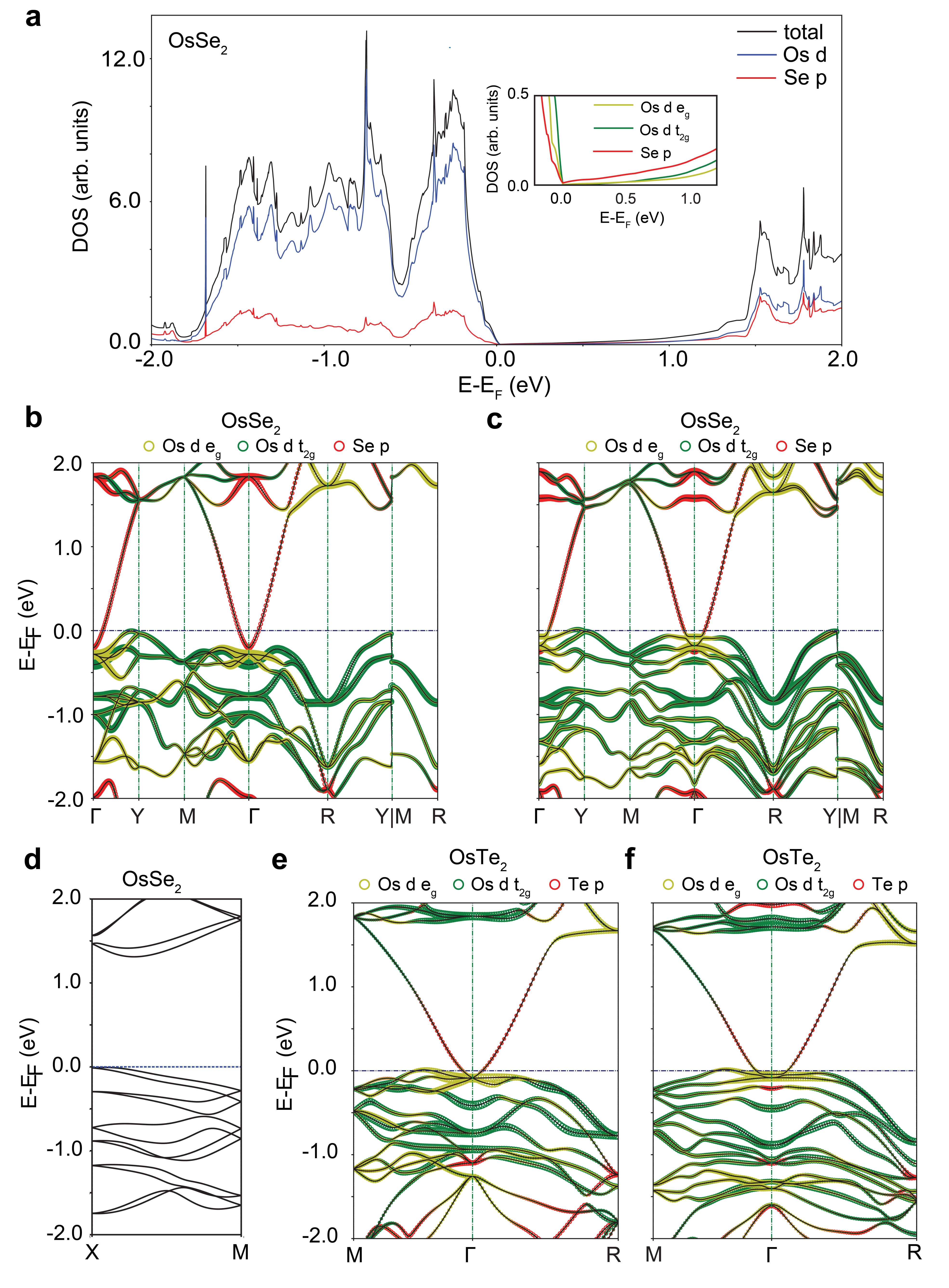}
 \end{center}
 \caption {\small (a) Total and projected electronic density of states of OsSe\textsubscript{2} in the absence of SOC, projected onto the Os $d$ orbitals and Se $p$ orbitals. (b-c) Band structure diagrams of OsSe\textsubscript{2}  without SOC (b) and with SOC (c). The projections of Os $d$ $e_{g}$, $d$ $t_{2g}$ and Se $p$ states are labelled in yellow, green and red circles, respectively. (d) Band dispersion of OsSe\textsubscript{2} along X--M direction with SOC. (e-f) Band structure diagrams of OsTe\textsubscript{2} (e) without SOC and (f) with SOC.} \label{fig2}
 \vspace{\baselineskip}
 \end{figure}
 \newpage
 \mbox{}
 \vfil
 \begin{figure}[htbp]
 \begin{center}
 \includegraphics[scale=0.23]{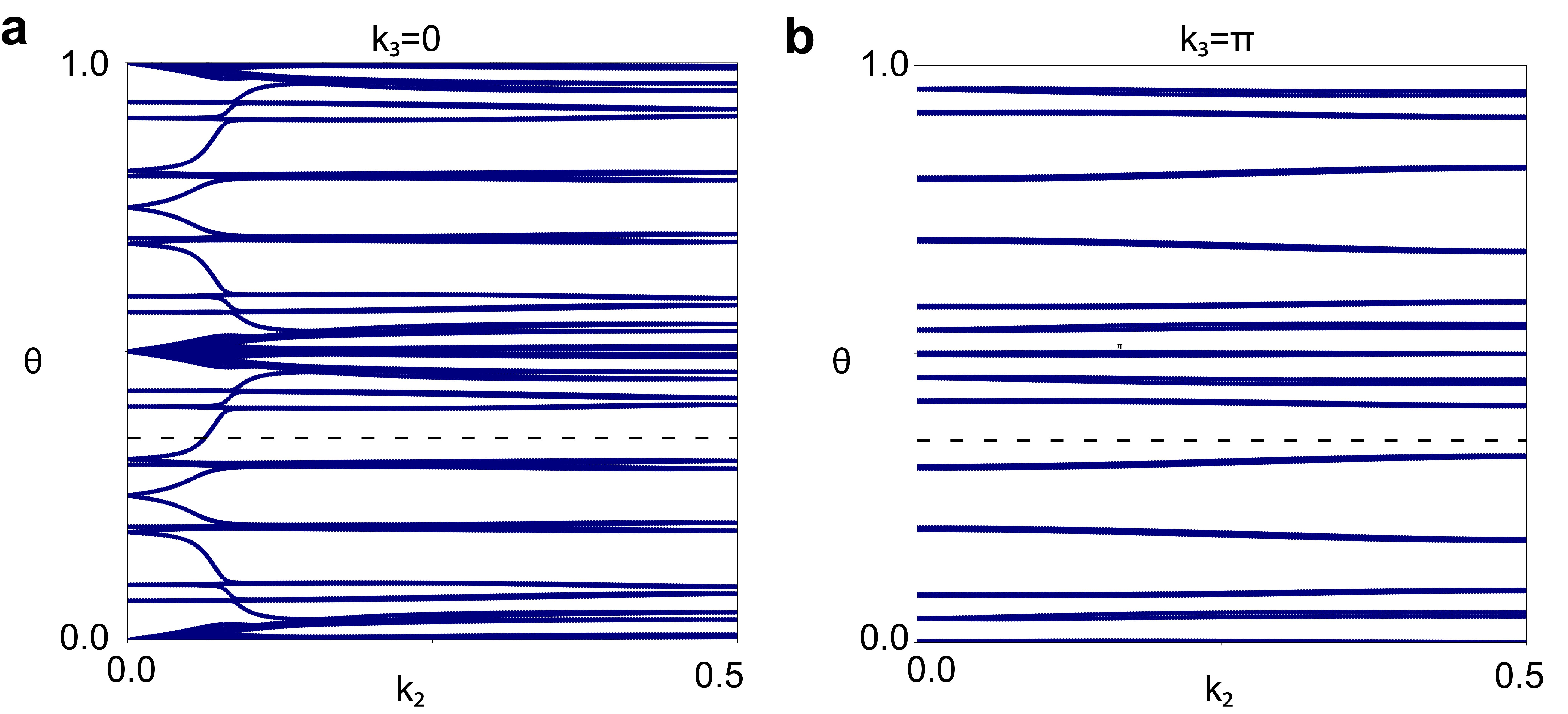}
 \end{center}
 \caption {\small The evolution of Wannier charge centers  along $k_{2}$ in (a) $k_{3}=0$ and (b) $k_{3}=\pi$ planes with SOC. The evolution line cross the reference line (dashed line) odd and even times in $k_{3}=0$ and $\pi$ planes, respectively.} \label{fig3}
 \vspace{\baselineskip}
 \end{figure}
 \vfill
 \mbox{}

 \begin{figure}[htbp]
 \begin{center}
 \includegraphics[scale=0.16]{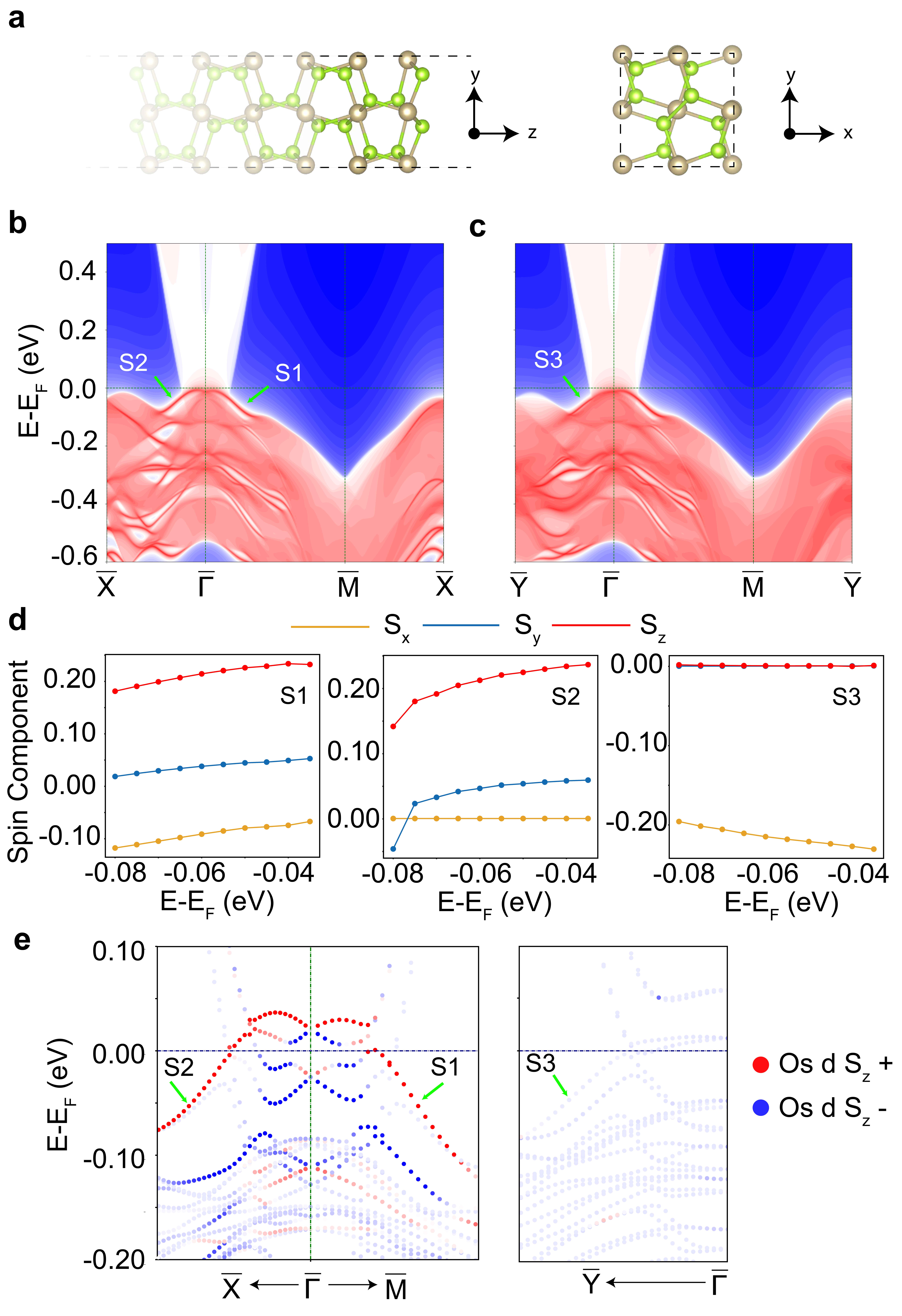}
 \end{center}
 \caption {\small (a) Side and top view of Se-terminated OsSe\textsubscript{2} (001)  surface. (b-c) Surface band diagrams for OsSe\textsubscript{2} obtained using non equilibrium Green's functions calculations. (d) Spin polarization along the surface bands S1, S2 and S3. (e) Corresponding surface band diagrams obtained from slab calculations.} \label{fig4}
 \vspace{\baselineskip}
 \end{figure}

 \begin{figure}[htbp]
 \begin{center}
 \includegraphics[scale=0.27]{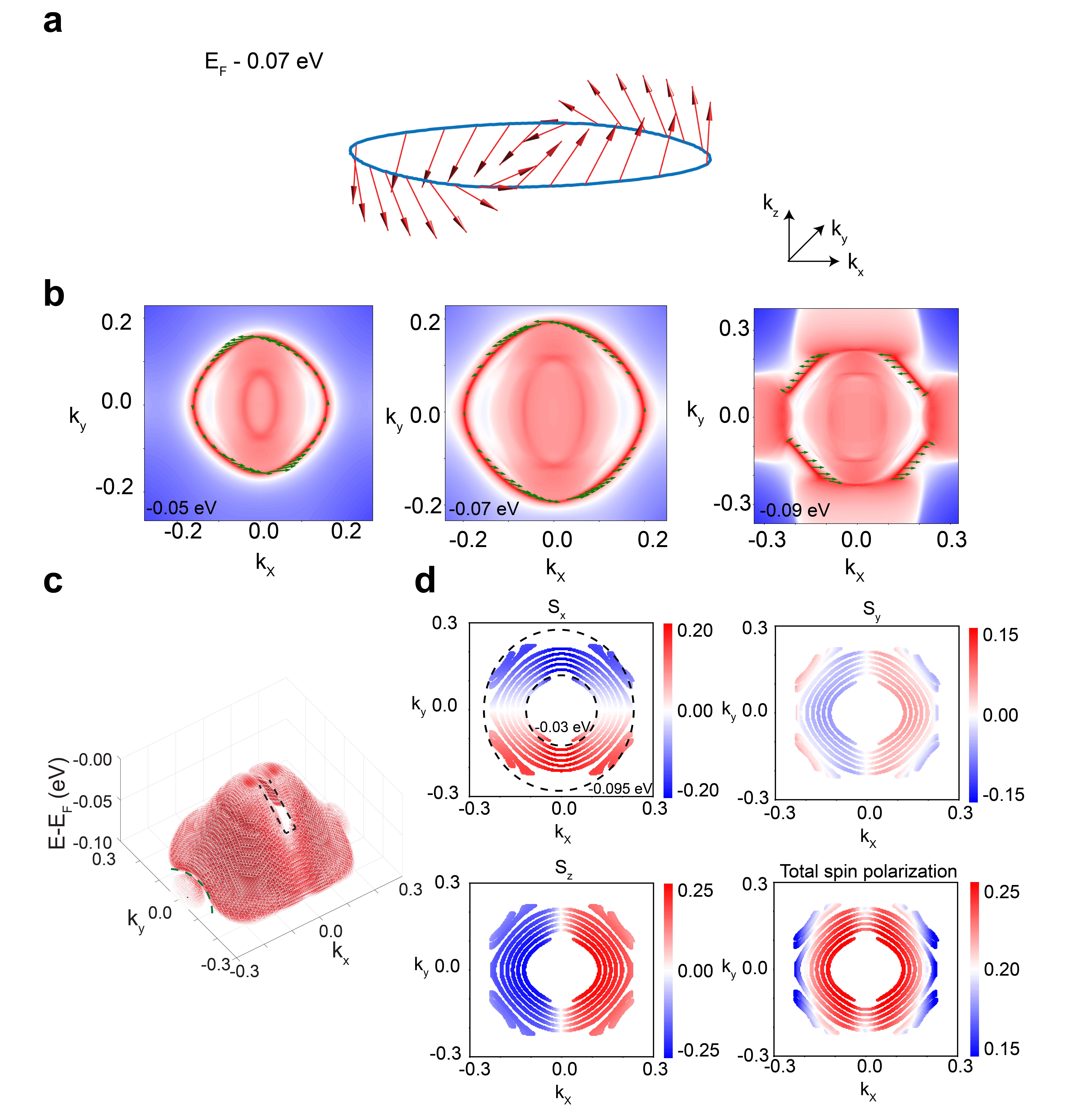}
 \end{center}
 \caption {\small Complex surface spin texture in OsSe\textsubscript{2}. (a) Three-dimensional vector representation of the surface spin texture at 0.07 eV below the Fermi level. (b) Spin textures at 0.05, 0.07 and 0.09 eV below the Fermi level.  (c) Surface band structure near the Fermi level ($E_{F}$-0.1 eV, $E_{F}$). The color difference represents the level of surface localization. (d) Evolution of surface spin polarization ($S_{x}$, $S_{y}$, $S_{z}$ and total spin polarization).
 } \label{fig5}
 \vspace{\baselineskip}
 \end{figure}

\end{document}